\documentclass[reprint,amsmath,amssymb,aps,prl,showpacs,superscriptaddress]{revtex4-1}
\usepackage{graphicx}

\begin{document}
\title{Experimental observation of polarization-dependent ultrafast carrier dynamics in multi-layer graphene}

\author{Xiao-Qing Yan}
\affiliation{%
The Key Laboratory of Weak Light Nonlinear Photonics, Ministry of Education, Teda Applied Physics School, and School of Physics, Nankai University, Tianjin 300457, China
}%
\affiliation{%
The Key Laboratory of Functional Polymer Materials and Center for Nanoscale Science \& Technology, Institute of Polymer Chemistry, College of Chemistry, Nankai University, Tianjin 300071, China
}%
\author{Zhi-Bo Liu}
\email{rainingstar@nankai.edu.cn}
\affiliation{%
The Key Laboratory of Weak Light Nonlinear Photonics, Ministry of Education, Teda Applied Physics School, and School of Physics, Nankai University, Tianjin 300457, China
}%
\author{Jun Yao}
\affiliation{%
The Key Laboratory of Weak Light Nonlinear Photonics, Ministry of Education, Teda Applied Physics School, and School of Physics, Nankai University, Tianjin 300457, China
}%
\author{Xin Zhao}
\affiliation{%
The Key Laboratory of Weak Light Nonlinear Photonics, Ministry of Education, Teda Applied Physics School, and School of Physics, Nankai University, Tianjin 300457, China
}%
\author{Xu-Dong Chen}
\affiliation{%
The Key Laboratory of Weak Light Nonlinear Photonics, Ministry of Education, Teda Applied Physics School, and School of Physics, Nankai University, Tianjin 300457, China
}%
\author{Xiang-Tian Kong}
\affiliation{%
The Key Laboratory of Weak Light Nonlinear Photonics, Ministry of Education, Teda Applied Physics School, and School of Physics, Nankai University, Tianjin 300457, China
}%
\author{Fei Xing}
\affiliation{%
The Key Laboratory of Weak Light Nonlinear Photonics, Ministry of Education, Teda Applied Physics School, and School of Physics, Nankai University, Tianjin 300457, China
}%
\author{Yongsheng Chen}
\affiliation{%
The Key Laboratory of Functional Polymer Materials and Center for Nanoscale Science \& Technology, Institute of Polymer Chemistry, College of Chemistry, Nankai University, Tianjin 300071, China
}%
\author{Jian-Guo Tian}
\email{jjtian@nankai.edu.cn}
\affiliation{%
The Key Laboratory of Weak Light Nonlinear Photonics, Ministry of Education, Teda Applied Physics School, and School of Physics, Nankai University, Tianjin 300457, China
}%

\date{\today}

\begin{abstract}
Polarization characteristic of ultrafast carrier dynamics in multi-layer CVD-grown graphene is probed with tilted beams (with respected to the graphene plane). The graphene ultrafast carrier dynamics measurement greatly depends on both polarization (i.e., orientation of linear polarization) and wave vector of probe beam. The differential reflectivity  $\Delta R/R$ signal of picosecond dynamics could be continuously altered from positive to negative by changing the probe polarization from P to S when the dynamics is probed by a total internal reflected beam. The polarization dependent  $\Delta R/R$  signal around 0 delay time is positive. It could be altered to negative by changing the probe polarization if the probe beam is non-total internal reflected beam. However, no sign reversal was observed for differential transmittance  $\Delta T/T$. These extremely unexpected results indicate the anisotropy of graphene carrier dynamics. Thus the ultrafast carrier dynamics should be further studied with consideration of the anisotropic structure (in- and out-of-graphene plane) of graphene.
\end{abstract}

\maketitle

Graphene, a single layer of carbon atoms in a honeycomb lattice, has aroused lots of works on its optical properties. Ultrafast carrier dynamics (UCD) of graphene and graphite thin films have been the subject of great interest due to their potential applications in photonic and electric devices \cite{Xia2009, Bonaccorso2010, Seibert1990, Breusing2011, Breusing2009, Sunphd, Sun2008, Malard2011, Ruzicka2012, Shang2010, Dawlaty2008, Wang2010, Winnerl2011, Carbone2011, Newson2009, Liprl2012, Huang2010, Butscher2007, Hale2011, Malic2011, Malic2012, Sun2012, Carbone2010}. At present, there are lots of works on the UCD in graphene and graphite \cite{Seibert1990, Breusing2011, Breusing2009, Sunphd, Sun2008, Malard2011, Ruzicka2012, Shang2010, Dawlaty2008, Wang2010, Winnerl2011, Carbone2011, Newson2009, Liprl2012, Huang2010, Butscher2007, Hale2011, Malic2011, Malic2012, Sun2012, Carbone2010}. The carrier dynamics could be characterized by two processes: thermalization and subsequent carrier cooling \cite{Malic2011}. Several physical mechanisms, including carrier-carrier scattering, carrier-optical photon scattering, carrier-acounstic phonon scattering and electron-hole recombination, have been advised for the photoexcited carrier relaxation \cite{Seibert1990, Breusing2011, Breusing2009, Sunphd, Sun2008, Malard2011, Ruzicka2012, Shang2010, Dawlaty2008, Wang2010, Winnerl2011, Carbone2011, Newson2009, Liprl2012, Huang2010}. Although there are extensive studies on the photoexcited carrier dynamics, the underlying physical mechanisms of the ultrafast nonlinear response in graphene still remain unclear.  The experimental results and the interpretations on the carrier dynamics are not consistent, e.g., negative  $\Delta T/T$ (positive  $\Delta R/R$ ) signal of picosecond (ps) dynamics (i.e., the so-called carrier cooling process with decay time in ps time scale) has been observed in  \cite{Seibert1990, Breusing2011, Breusing2009, Sunphd, Sun2008, Malard2011, Ruzicka2012, Shang2010}. However, it was positive for  $\Delta T/T$ (negative for  $\Delta R/R$ ) in \cite{Dawlaty2008, Wang2010, Winnerl2011, Carbone2011, Newson2009, Liprl2012}. The negative $\Delta R/R$ was ascribed to doping effect by Sun \textit{et al} \cite{Sun2008}, Breusing \textit{et al} attribute it to the renormalization of electron and hole states by transient carrier population \cite{Breusing2011}, Carbone  regards the valley of $\Delta T$ is induced by the interlayer interaction \cite{Carbone2010}, and so on. A reasonable explanation for this occasionally-observed negative  $\Delta T/T$ signal and positive $\Delta R/R$ signal is expected. In addition, the reason for the non-uniform measured decay time of ps dynamics is not unknown \cite{Breusing2009,Sun2008}.

The $\Delta R/R$ ($\Delta T/T$) around 0 ps delay time (i.e., $(\Delta R/R)|_\text{0}$ and $(\Delta T/T)|_\text{0}$ ) was negative (positive) in literature \cite{Seibert1990, Breusing2011, Breusing2009, Sunphd, Sun2008, Malard2011, Ruzicka2012, Shang2010, Dawlaty2008, Wang2010, Winnerl2011, Carbone2011, Newson2009, Liprl2012}. The initial  increase of the  $\Delta T/T$ signal is due to the generation of carriers during the applied pulse, the increased carrier occupation leads to an absorption bleaching \cite{Seibert1990, Breusing2011, Breusing2009, Sunphd, Sun2008, Malard2011}. Thus positive $(\Delta R/R)|_\text{0}$ should be reasonable \cite{Seibert1990}. The carrier relaxation processes were the focus in previous works, the relationship between probe absorption blocking and polarization of beams was seldom involved.

Polarized lights had been used to study the excited spin polarized carrier dynamics from  $\Delta T/T$, however, no polarization dependence of $\Delta T/T$ was observed \cite{Sunphd} . Indeed, we have not observed any visible polarization dependent pump-probe signal in degenerate  $\Delta T/T$ measurements with 800 nm probe beam nearly normal to the graphene plane (not shown here). Malic \textit{et al} have theoretically shown that the photoexcited carrier density distribution already during the pulse excitation was highly anisotropic and the carrier distribution could become completely isotropic already after the first 50 femtoseconds (fs) \cite{Malic2011, Malic2012}. Therefore, the degenerate pump-probe  $\Delta T/T$ signal around the first tens of fs delay time should be polarization dependent and the  $\Delta T/T$ signal of ps dynamics does not change with polarization of beams. Until now, the polarization dependence of $\Delta T/T$ (or $\Delta R/R$) signal has not been reported in graphene.

Previous studies were conducted with beams (nearly) normal to the graphene plane \cite{Breusing2011, Breusing2009, Sunphd, Sun2008,Malard2011, Ruzicka2012, Shang2010, Dawlaty2008, Wang2010, Winnerl2011, Carbone2011, Newson2009, Liprl2012, Huang2010}. The probe beam monitors the carrier at state of $\pm \hbar \omega_\text{probe}/2$. The non-equilibrium electron dynamics was regarded to be similar to that of hole in conduction band due to the identical valence and conduction band structure \cite{Breusing2011, Breusing2009, Sunphd, Sun2008, Malard2011, Ruzicka2012, Shang2010, Dawlaty2008, Wang2010, Winnerl2011, Carbone2011, Newson2009, Liprl2012, Huang2010, Butscher2007, Hale2011, Malic2011, Malic2012}. If the probe (pump) beam is incident obliquely on graphene, the probe (pump) beam may no longer monitors (excites) carriers at state of $\pm \hbar \omega_\text{probe}/2$ ($\pm \hbar \omega_\text{pump}/2$) \cite{Makprl2008}. The pump-probe signal may be different. In the other hand, the carrier motion is anisotropic for in-graphene plane and out-of-graphene plane due to the two-dimensional structure of graphene, the graphene carrier dynamics could be anisotropic.

In this letter, we study the polarization dependence of UCD in graphene by using tilted beams. The pump-probe signal, including  $\Delta R/R$  and  $\Delta T/T$, greatly depends on both polarization and wave vector of beams. When UCD was probed with a total internal reflected beam, the $(\Delta R/R)|_\text{0}$ is positive and could be modified by adjusting polarizations of probe and pump beams. Most remarkably, the  $\Delta R/R$  signal of ps dynamics could be altered from positive to negative by changing the probe polarization from P to S. When UCD was probed with a non total internal reflected beam, the $(\Delta R/R)|_\text{0}$ could be altered from positive to negative by changing probe polarizations from S to P. However, the  $\Delta T/T$ changes with polarization of beams and keeps positive. Our measurements extend previous studies by considering wave vector and polarization of beams. These unexpected results call for a systematical theoretical explanation for the anisotropy of carrier dynamics. The answer to the occasionally-observed negative  $\Delta T/T$ and non-uniform decay time of ps dynamics may be obtained during the theoretical analysis. Observing the anisotropy of carrier dynamics and absorption blocking is an important strength of our approach.

\begin{figure}
  \includegraphics[width=\columnwidth]{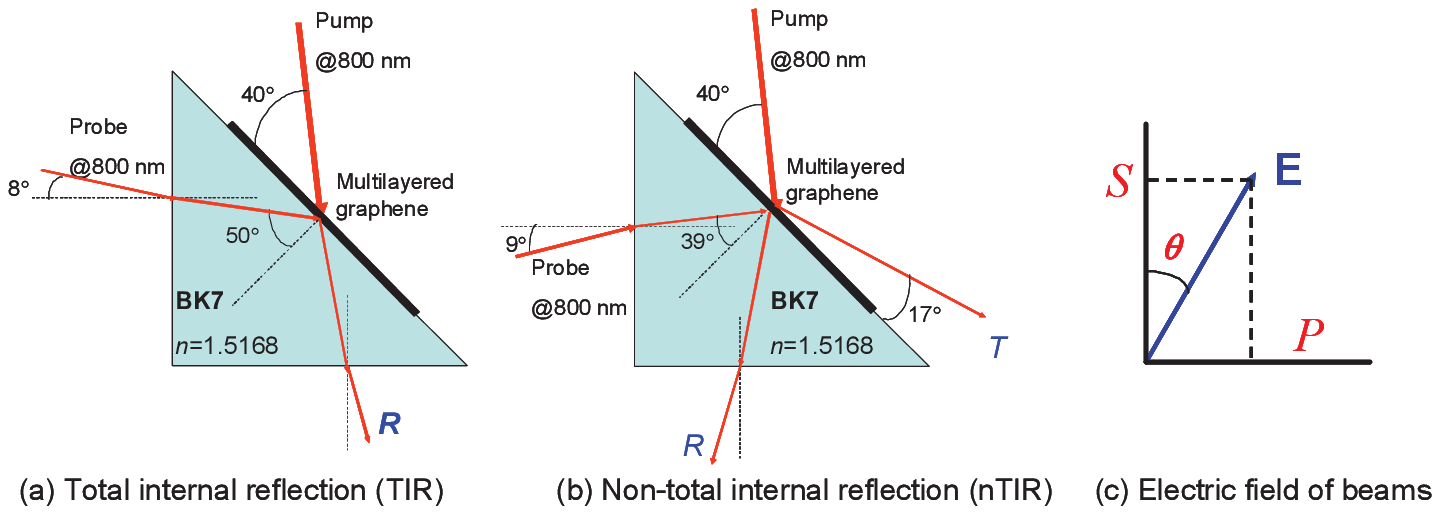}\\
  \caption{(color online) Experimental setups (a,b) and definition (c) of polarization of beams.
  }\label{Fig1}
\end{figure}

The CVD-grown multi-layer graphene was purchased from Graphene-Supermarket \cite{Graphene1} and subsequently transferred to the inclined plane of a BK 7 right-angle prism \cite{Chen2013}. The layer number has been checked to be 3-5 from Raman spectroscopy \cite{Xing2012}.
Our experimental setups are shown in Fig.~\ref{Fig1}. For the case of total internal reflection (TIR), the reflectance change was measured. For the case of non-total internal reflection (nTIR), both the reflectance and transmittance changes were measured. The polarizations of pump and probe beams is defined as Fig.~\ref{Fig1}c shown.

The laser source  was a Ti: sapphire regenerative amplifier (Spitfire Pro, Spectra Physics) and was operated at the wavelength of 800 nm (1 kHz, 120 fs (FWHM)). The 800-nm laser pulses served as both pump and probe light. Two $\lambda/2$ plate-Glan Talyor prism-$\lambda/2$ plate combinations were inserted into the probe and pump beams to alter the pulse energy and polarization. The pump and probe beams were focused at the sample by a 400 and 200 mm focal-length lens, respectively. The measurement system is similar to that in \cite{Zhaoapl2011}. After passing the optical components before sample, the pulse duration was about 320 fs determined from transient two-photon absorption measurement of ZnSe \cite{Krauss1994}. Since the pulse width is much larger than the time scale of thermalization of non-equilibrium carrier distribution \cite{Breusing2011,Breusing2009,Malic2011}, the discussions on decay time have to be limited to the process of ps dynamics. All these measurements were performed at ultraclean room ($T=20~^\circ$C), the effect of lattice temperature on graphene carrier dynamics is identical in our pump-probe measurements \cite{Sun2008}. Thus the observed polarization dependence of signal is unrelated to the lattice temperature. Measurements with different graphene samples give similar results although the CVD-grown graphene is not as homogeneous as exfoliated graphene \cite{Ruzicka2012}. The stack order, layer number and defect could not be the origin of the observed polarization dependence. The observed polarization dependence should be from intrinsic optical response of graphene. In the following, the experimental results from the two setups are presented separately.

For both TIR and nTIR, the linear reflectance and transmittance of probe beam depend on the polarization (see Fig. S1 in supplemental material \cite{SM1}). Polarization dependence of linear reflectance and transmittance could be well explained with the theory based on the boundary continuity of Maxwell equitation \cite{Xing2012, Yeapl2013}.

\begin{figure}
  \includegraphics[width=0.8\columnwidth]{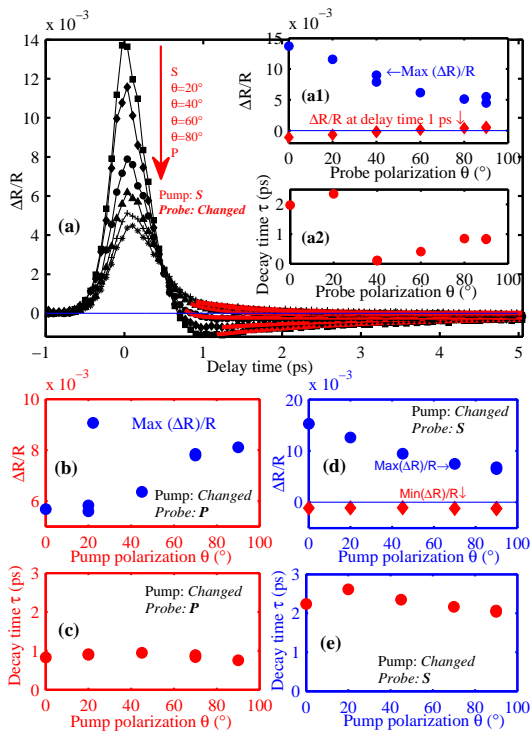}\\
  \caption{(color online) Polarization dependent $\Delta R/R$ for TIR. (a) The $\Delta R/R$ for probe beam with different polarization under pump fluence of 0.16 mJ/cm$^2$, the probe polarization dependences of the peak  $\Delta R/R$ (a1),  $\Delta R/R$ value at 1 ps delay time (a1) and decay time of ps dynamics (a2) are summarized in the insert. The pump polarization dependence of max($\Delta R/R$) (b) and decay time of ps dynamics (c) when the probe beam is \textit{p-}polarized. The max($\Delta R/R$) (d) and decay time of ps dynamics (f) as function of pump polarization when the probe beam is \textit{s-}polarized.
  }\label{Fig2}
\end{figure}

Firstly, we focus on the experimental results of TIR. Figure \ref{Fig2}a shows representative  $\Delta R/R$  time scan for different polarized probe beam. The  $\Delta R/R$  signal immediately following the pump is positive and peaks around 0 delay time. It greatly depends on the probe polarization. As expected, positive  $\Delta R/R$  signal is observed during pulse excitation. After pulse excitation, the positive  $\Delta R/R$  signal decreases towards zero due to carrier relaxation. The  $\Delta R/R$  becomes negative at a delay time and subsequently relaxes to 0. The delay time of  $\Delta R/R$  sign reversing increases monotonously with altering probe polarization from S to P. The maximum  $(\Delta R/R)|_\text{0}$ (i.e., max($\Delta R/R$)) and the  $\Delta R/R$  at 1 ps delay time are summarized in Fig.~\ref{Fig2}a1. We see that the max( $\Delta R/R$ ) decreases from \textit{s-}polarized to \textit{p-}polarized light, the  $\Delta R/R$  at 1 ps delay time increases from negative to positive when probe polarization changes from S to P. We used a single exponential function to fit the  $\Delta R/R$  at long delay time to obtain the decay time for ps dynamics, the extracted decay time is summarized in Fig.~\ref{Fig2}a2. The decay time greatly depends on probe polarization, and the decay time measured by \textit{p-}polarized probe beam agrees with that in \cite{Breusing2011, Breusing2009}.

Similar probe polarization dependence of  $\Delta R/R$ was observed for \textit{p-}polarized pump beam, except that the max($\Delta R/R$) is largest for \textit{p-}polarized probe beam and smallest for \textit{s-}polarized probe beam (see Fig. S2 in supplemental material \cite{SM1}).

Sun \textit{et al} has observed a similar continuous change in $\Delta T/T$ signal of ps dynamics from negative at low lattice temperature to positive at high lattice temperature \cite{Sun2008}. In our measurements, the observed sign reversal of $\Delta R/R$ signal is with probe polarization. So, the signal of ps dynamics is related to both the lattice temperature and polarization of probe beam.

Figures \ref{Fig2}b and d summarize the pump polarization dependence of max($\Delta R/R$) and the minimum  $\Delta R/R$  around 1 ps delay time (i.e., min($\Delta R/R$)) for both \textit{p-} and \textit{s-}polarized probe beams. For \textit{p-}polarized probe beam, the  $\Delta R/R$  is largest for \textit{p-}polarized light excitation. However, if the probe beam is \textit{s-}polarized, the  $\Delta R/R$  is largest when the pump beam is \textit{s-}polarized. In addition, the valley value of  $\Delta R/R$ and decay time of ps dynamics is  pump polarization independent (Fig.~\ref{Fig2}c and e). Clearly, the  $\Delta R/R$  curve could be divided into two parts. One part is pump polarization dependent and another part is pump polarization independent. Corresponding, the carrier dynamics in the two parts is different. The positive pump-probe signal, i.e., pump-induced reflectivity here, is caused by the bleaching of interband transition due to state filling effect \cite{Seibert1990}. The initial pump polarization dependence of $\Delta R/R$ orignates from the anisotropic non-equilibrium carrier distribution already after pulse excitation \cite{Malic2012}. Since the anisotropic carrier distribution could become completely isotropic within 50 fs after pulse excitation, the  $\Delta R/R$  signal of ps dynamics is pump polarization independent.

The peak  $\Delta R/R$  value is linear dependent on pump fluence no matter what the polarizations of probe and pump beams. However, the valley  $\Delta R/R$  value decreases with pump fluence at low pump energy and does not change at high pump fluence (see Fig. S3 in supplemental material \cite{SM1}).

\begin{figure}
  \includegraphics[width=0.8\columnwidth]{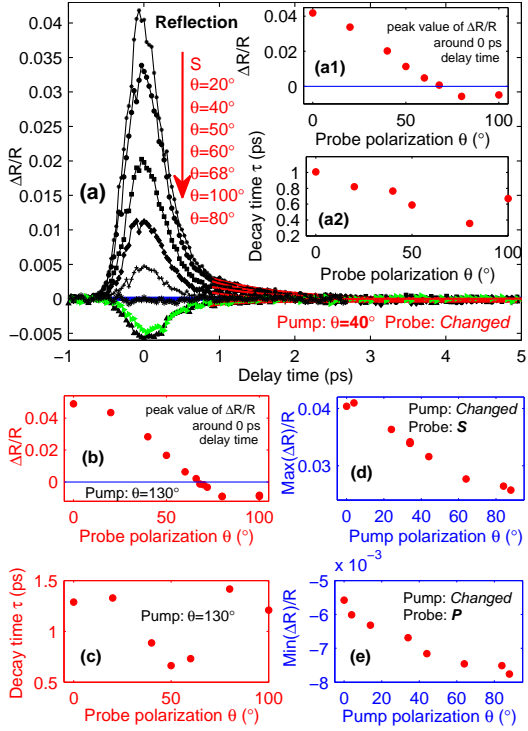}\\
  \caption{(color online) Polarization dependent $\Delta R/R$ for nTIR. (a) The $\Delta R/R$ for different polarized probe beam under pump fluence of 0.2 mJ/cm$^2$, the probe polarization dependences of the peak/valley $(\Delta R/R)|_\text{0}$ (a1) and decay time of ps dynamics (a2) are summarized in the insert. The probe polarization dependence of peak $(\Delta R/R)|_\text{0}$ (b) and decay time of ps dynamics (c) when the pump polarization is  $\theta=130^{\circ}$. The peak/valley $(\Delta R/R)|_\text{0}$ as function of pump polarization for $s$-polarized (d) and $p$-polarized (e) probe beam.
  }\label{Fig3}
\end{figure}

Now, we turn to the experiments probed by non-total reflected light. Figure 3a presents transient  $\Delta R/R$  time scan for probe beam with different polarization when the pump polarization is $\theta=40^{\circ}$. Similar experimental results has been observed under the excitation of pump beam with $\theta=130^{\circ}$ (see Fig. S4 in supplemental material \cite{SM1}). The $(\Delta R/R)|_\text{0}$ is positive for \textit{s-}polarized probe beam and it could be continuously altered to negative by changing the probe polarization from S to P. In addition, there is a polarization of $\theta=68^{\circ}$ for the probe beam without reflectance change no matter what the pump polarization is (Figs.~\ref{Fig3}a and b). A sign reversal of the pump-probe  $\Delta T/T$ signal around 0 delay time has observed as photon energies are varied across the Fermi level by Winnerl \textit{et al} \cite{Winnerl2011}. This sign reversal is induced by the interplay between inter- and intraband absorption of photon. The Fermi energy of graphene does not change with probe polarization in our measurements. The sign reversal of $(\Delta R/R)|_\text{0}$ with probe polarization is different from that in \cite{Winnerl2011}.

As the case of TIR, both the peak and valley value of $(\Delta R/R)|_\text{0}$ is linearly dependent on pump fluence (see Fig. S5 in supplemental material \cite{SM1}). Mikhailov and Ziegler shows that \textit{s-}polarized light can propagate in the graphene plane if the photon energy of probe beam falls into the window of  $1.667\mu<\hbar\omega<2\mu$ ($\mu$: chemical potential) \cite{Mikhailov2007}. The negative  $\Delta R/R$  signal could not be from the leakage of \textit{s-}polarized light due to $(\Delta R/R)|_\text{0}$ is positive for \textit{s-}polarized probe beam for both TIR and nTIR. And, the linear pump fluence dependence of max($\Delta R/R$) indicates that the  $\Delta R/R$  signal is from blocking of single photon absorption \cite{Seibert1990}.

The probe polarization dependence of peak/valley $(\Delta R/R)|_\text{0}$ is similar for the two types of pump beam (Fig.~\ref{Fig3}a1 and b). The decay time depends on probe polarization, but the dependence is not as strong as that at the case of TIR (Figs.~\ref{Fig2}a2,~\ref{Fig3}a2 and~\ref{Fig3}c).

As TIR, pump polarization could also determine the reflectance change, the max($\Delta R/R$) for \textit{s-}polarized probe beam is largest for \textit{s-}polarized pump beam and smallest for \textit{p-}polarized pump beam (Fig.~\ref{Fig3}d). For \textit{p-}polarized probe beam (Fig.~\ref{Fig3}e), the negative minimum $(\Delta R/R)|_\text{0}$ is largest for \textit{p-}polarized pump beam and smallest for \textit{s-}polarized pump beam.

\begin{figure}
  \includegraphics[width=0.7\columnwidth]{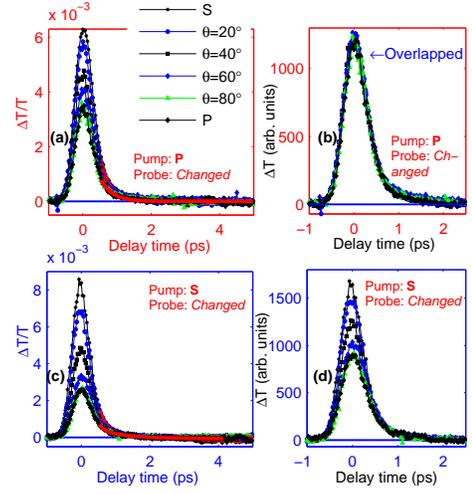}\\
  \caption{(color online) Polarization dependent  $\Delta T/T$ and $\Delta T$ under pump fluence of 0.2 mJ/cm$^2$ for nTIR. The $\Delta T/T$ (a) and transmittance change $\Delta T$ (b) for different polarized probe beam  when the pump beam is \textit{p-}polarized. The $\Delta T/T$ (c) and $\Delta T$ (d) for different polarized probe beam when the pump beam is \textit{s-}polarized.
  }\label{Fig4}
\end{figure}

For nTIR, we have not observed the sign reversal in $\Delta R/R$  signal of ps dynamics. Comparing the experimental results in Fig.~\ref{Fig3} and Fig.~\ref{Fig4}, we see that both the $(\Delta R/R)|_\text{0}$ signal and the  $\Delta R/R$  signal of ps dynamics are related to the wave vector of probe beam.

To compare with the reflectance change, the transient transmittance change was also measured for the case of nTIR. Figure \ref{Fig4}a and c show the differential transmittance $\Delta T/T$ time scans for different polarized probe beam under \textit{p-} and \textit{s-}polarized light excitation, respectively. For the two types of pump beam, the  $\Delta T/T$ decreases with probe polarization when the probe polarization is changed from S to P. The decay time of ps dynamics is extracted to be $\sim0.9$ ps, which is in good agreement with the decay time measured with \textit{p-}polarized probe beam in TIR. However, not as TIR, the decay time is independent on polarizations of both pump and probe beams.

If the pump beam is \textit{p-}polarized the transmittance change $\Delta T$ does not change with probe polarization (Fig.~\ref{Fig4}b). But, the $\Delta T$ depends on probe polarization if the pump beam is \textit{s-} polarized. For \textit{p-}polarized pump beam, the probe polarization dependence of  $\Delta T/T$ is caused by the change of transmittance \textit{T} with probe polarization (see Fig. S1c of supplemental material in \cite{SM1}). It is noted that all the reflectance change $\Delta R$ shown above is probe polarization dependent. Comparing the polarization dependence of the two types of pump beam, we can conclude that the probe polarization dependence of $\Delta T$ is eliminated if the electric field of pump beam is not in graphene plane.

As the case of reflection, the max($\Delta T/T$) depends on the polarizations of both probe (Fig.~\ref{Fig4}) and pump beams and the max($\Delta T/T$) is linearly dependent on pump fluence (see Figs. S6 and S7 of supplemental material in \cite{SM1}). For \textit{s-}polarized probe beam, the max($\Delta T/T$) decreases with altering pump polarization from S to P. If the probe beam is \textit{p-}polarized, the max($(\Delta T/T)$) signal increases with altering pump polarization from S to P.

We now turn to the polarization dependence of $(\Delta R/R)|_\text{0}$ and $(\Delta T/T)|_\text{0}$. The changing of polarization and wave vector just adjusts the electric field component along in-graphene plane and out-of graphene plane. Thus the pump-induced reflectance and transmittance change is different for $E_{\text{in-graphene}}$ and $E_{\text{out-of-graphene}}$. Taking into account the linear pump fluence dependence of signal, the influence of pump beam on the electric field components of probe beam could be written as
\begin{align}\label{eqt1}
\Delta E_{i,\text{probe}}^{R} / E_{i,\text{probe}}^{R} = \sum_j a_{ij}(\omega_{\text{probe}}, \omega_{\text{pump}}, t) E_{j,\text{pump}} \nonumber \\
\Delta E_{i,\text{probe}}^{T} / E_{i,\text{probe}}^{T} = \sum_j b_{ij}(\omega_{\text{probe}}, \omega_{\text{pump}}, t) E_{j,\text{pump}}
\end{align}
where $i,j=x,y,z$ (\textit{x},\text{y}-in graphene plane, $a_{ij}$ and $b_{ij}$ are second-rank tensors used to describe the pump-induced absorption change for reflectance and transmittance, respectively.
The influence of carrier relaxation on $a_{ij}$ and $b_{ij}$ is through the delay time \textit{t} . The polarization independence of carrier dynamics for normal incident probe beam indicates the \textit{x-y} plane symmetry for graphene, thus the coordinate axes of $x$ and $y$ must be equivalent. It is clear that these tensors possess the following symmetry properties: $ a_{xx}=a_{yy},a_{xy}=a_{yx},a_{xz}=a_{yz},a_{zx}=a_{zy} $ and $ b_{xx}=b_{yy},b_{xy}=b_{yx},b_{xz}=b_{yz},b_{zx}=b_{zy}$.
As shown in Eq.~\ref{eqt1}, the change of $E_{i,\text{probe}}$ in one direction is related to all the electric field components of pump beam through  $a_{ij}$ and  $b_{ij}$ . Owing to both wave vector and polarization could alter the electric field component along each direction, the pump induced reflectivity and transmittance change is polarization and wave vector dependent. The polarization dependence could eliminate gradually with the photoexcited carrier relaxation.

For the polarization dependence of ps dynamics, the physical mechanism is not known. And, it seems that the reported interpretations for the negative signal of $\Delta T/T$ could not well work for this polarization dependence \cite{Seibert1990, Breusing2011, Breusing2009, Sunphd, Sun2008, Malard2011, Ruzicka2012, Shang2010}. More theoretical studies should be devoted to get a clear comprehension of the photoexcited carrier relaxation.

In summary, the graphene ultrafast carrier dynamics measurements depend on both the polarization of beams and the wave vector of probe beam. The carrier relaxation and photon absorption blocking in graphene are anisotropic. The systematical physical mechanisms for the photon absorption blocking and, in particular, ps dynamics taking into account the anisotropy of graphene are being called to complement the experiments. In addition, our findings suggest that the optical modulation in graphene is polarization and wave vector controllable.

\begin{acknowledgments}
This work was supported by the Chinese National Key Basic Research Special Fund (2011CB922003), the National Natural Science Foundation of China (11174159), and the China Postdoctoral Science Foundation funded project (012M510743).
\end{acknowledgments}

\end{document}